\documentclass[11pt,a4paper]{article}

\usepackage{amssymb}
\usepackage[dvips]{graphicx}
\usepackage{bm}

\begin{document}

\title{On the two-loop divergences of the 2-point hypermultiplet supergraphs for $6D$, ${\cal N} = (1,1)$ SYM theory}

\author{I.L. Buchbinder\footnote{joseph@tspu.edu.ru}\\
{\small{\em Department of Theoretical Physics, Tomsk State Pedagogical
University,}}\\
{\small{\em 634061, Tomsk,  Russia}} \\
{\small{\em National Research Tomsk State University, 634050, Tomsk, Russia}},\\
\\
E.A. Ivanov\footnote{eivanov@theor.jinr.ru}\\
{\small{\em Bogoliubov Laboratory of Theoretical Physics, JINR, 141980 Dubna, Moscow region,
Russia}},\\
\\
B.S. Merzlikin\footnote{merzlikin@tspu.edu.ru}\\
{\small{\em Department of Theoretical Physics, Tomsk State Pedagogical
University}},\\
{\small{\em 634061, Tomsk,  Russia}}, \\
{\small{\em Department of Higher Mathematics and Mathematical Physics}},\\
{\small{\em \it Tomsk Polytechnic University, 634050, Tomsk, Russia}},\\
\\
K.V. Stepanyantz\footnote{stepan@m9com.ru}\\ {\small{\em Moscow State University}},\\
{\small{\em Faculty of Physics, Department of Theoretical Physics}},\\
{\small{\em 119991, Moscow, Russia}} }

\date{}

\maketitle

\begin{abstract}
We consider $6D$, ${\cal N}=(1,1)$ supersymmetric Yang-Mills theory
formulated in ${\cal N}=(1,0)$ harmonic superspace and analyze  the
structure of the two-loop divergences in the hypermultiplet sector.
Using the ${\cal N}=(1,0)$ superfield background field method we
study the two-point supergraphs with the hypermultiplet legs and
prove that their total contribution to the divergent part of
effective action vanishes off shell.
\end{abstract}

\unitlength=1cm

\section{Introduction}
\hspace*{\parindent}

This paper is a continuation and further development of our previous
works on the structure of divergences in $6D$, ${\cal N}=(1,0)$ and
${\cal N}=(1,1)$ gauge theories
\cite{Buchbinder:2016gmc,Buchbinder:2016url,Buchbinder:2017ozh}.

The study of supersymmetric gauge models in higher dimensions
attracts much attention due to  both their tight links with
the superstring/brane stuff and some remarkable properties of them
in the quantum domain. On the one hand, these models are
non-renormalizable because of the dimensionful coupling constant,
like, e.g., (super)gravity theories. On the other hand, supersymmetry
may ensure canceling some divergences, so that one can expect a
better ultraviolet behavior of supersymmetric gauge theories as compared to the
non-supersymmetric case
\cite{Howe:1983jm,Howe:2002ui,Bossard:2009sy,Bossard:2009mn,Fradkin:1982kf,Marcus:1983bd,Marcus:1984ei,Bork:2015zaa,Bossard:2015dva}
(see also the review \cite{Smilga:2016dpe}).

It is known that in $4D, \,{\cal N}=4$ SYM theory  all
divergences vanish off-shell and the theory proves to be finite just due to a
large amount of supersymmetries \cite{Grisaru:1982zh,Mandelstam:1982cb,Brink:1982pd,Howe:1983sr}. At the classical level,
this theory is very similar to $6D, \,{\cal N}=(1,1)$ SYM theory (see, e.g., \cite{Howe:1983fr} for a formulation of $6D$ supersymmetry). Indeed,
$4D, \,{\cal N}=4$ SYM theory can be obtained from $6D \,{\cal N}=(1,1)$ theory by means of dimensional
reduction. Moreover, formulations of both theories in the
harmonic superspace \cite{Howe:1985ar,Zupnik:1986da,Ivanov:2005qf,Ivanov:2005kz,Buchbinder:2014sna,Bossard:2015dva} reveal a great similarity.
This resemblance suggests that $6D, \, {\cal N}=(1,1)$ SYM theory
could have a better ultraviolet behavior compared to other $6D$
theories. This was confirmed by the one-loop calculation, which
demonstrated that  one-loop divergences in this theory cancel even off-shell \cite{Buchbinder:2016url,Buchbinder:2017ozh}. Obviously, it would be very interesting to
investigate whether this remarkable quantum property persists at the two-loop level. It is known that
$6D, \,{\cal N}=(1,1)$ SYM theory is on-shell finite at the two-loop
level \cite{Howe:1983jm,Howe:2002ui,Bossard:2009sy,Bossard:2009mn,Fradkin:1982kf,Marcus:1983bd,Bork:2015zaa}. In this letter we will investigate the
off-shell divergences. To calculate them,  we make use of the technique of
the harmonic supergraphs, which allows one to perform all calculations in a manifestly
${\cal N}=(1,0)$ supersymmetric way. Besides, we use the ${\cal
N}=(1,0)$ background superfield method which ensures preserving the classical gauge
symmetry of the effective action \cite{Buchbinder:2016url,Buchbinder:2017ozh}. The theory under consideration possesses the
hidden  ${\cal N}=(0,1)$ supersymmetry \cite{Bossard:2015dva}. As a consequence,
it suffices to analyze  the supergraphs with the hypermultiplet external
lines only. All other supergraphs can be obtained from these ones via
the hidden supersymmetry transformations. Therefore, if the
effective action  is finite in the pure hypermultiplet sector, it is finite as a total. In this letter we limit our study to the
structure of the two-loop divergences of the supergraphs with the two
external hypermultiplet legs only and demonstrate that all such divergences cancel off shell. The divergences of the supergraphs
with four hypermultiplet  legs will be a subject of the next publication.

\section{${\cal N}=(1,1)$ SYM theory in ${\cal N}=(1,0)$ harmonic superspace}
\hspace*{\parindent}

The $6D, \,{\cal N}=(1,1)$ SYM theory can be considered as a particular
case of $6D, \,{\cal N}=(1,0)$ SYM theories. It is convenient to
describe them using the formalism of the harmonic superspace,
because ${\cal N}=(1,0)$ supersymmetry is then a manifest symmetry
of the theory at all steps of calculating quantum corrections.
Besides, this theory possesses the hidden ${\cal N}=(0,1)$
supersymmetry \cite{Bossard:2015dva}.

Following ref. \cite{Bossard:2015dva} we briefly consider the harmonic superspace formulation of the
theory. We introduce the harmonic
variables $u^{\pm i}$, where $i=1,2$ and $u_i^- = (u^{+i})^*$, such that $u^{+i}
u_i^- = 1$. The harmonic superspace is obtained by adding these
coordinates to the set $z\equiv (x^M,\theta^a_i)$, where $x^M$ with
$M=0,\ldots, 5$, are usual $6D$ Minkowski coordinates and $\theta^{a}_i$ with $a=1,\ldots,4$ are
anti-commuting left-handed spinors. The analytic coordinates are
defined as $\zeta\equiv (x^M_A,\theta^{+a})$, where

\begin{equation}
x^M_A \equiv x^M + \frac{i}{2}\theta^{-}\gamma^M \theta^+;\qquad \theta^{\pm a} \equiv u^\pm_i \theta^{ai},
\end{equation}

\noindent
with $\gamma^M$ being the six-dimensional $\gamma$-matrices. In our notation, the integration measures are written as

\begin{equation}\label{Integrations}
\int d^{14}z = \int d^6x\,d^8\theta;\qquad \int d\zeta^{(-4)} \equiv \int d^6x\, d^4\theta^+.
\end{equation}

In the harmonic superspace approach, $6D, {\cal N}=(1,0)$ SYM theory with the gauge group $G$ and the hypermultiplets
in the representation $R$ is described by the action

\begin{eqnarray}\label{Action}
S = \frac{1}{f_0^2} \sum\limits_{n=2}^\infty \frac{(-i)^{n}}{n} \mbox{tr} \int d^{14}z\, du_1 \ldots du_n\,
\frac{V^{++}(z,u_1)\ldots V^{++}(z,u_n)}{(u_1^+ u_2^+) \ldots (u_n^+ u_1^+)} - \int d\zeta^{(-4)} du\,\widetilde q^+ \nabla^{++} q^+.
\end{eqnarray}

\noindent Then ${\cal N}=(1,1)$ SYM theory is reproduced in the
particular case, when the hypermultiplets belong to the adjoint
representation, $R=Adj$. In general, the theories described by the action (\ref{Action}) are anomalous \cite{Townsend:1983ana,Smilga:2006ax,Kuzenko:2015xiz},
but the anomalies are canceled for ${\cal N}=(1,1)$ theory.

The gauge superfield $V^{++}(z,u)$ lies in the adjoint
representation of the gauge group. It is real with respect to the
specially defined conjugation denoted by tilde, $\,\widetilde{V^{++}} = V^{++}\,$, and satisfies the analyticity
condition,

\begin{equation}
D^+_a V^{++} = 0.
\end{equation}

\noindent
In the pure gauge field part of the action (\ref{Action}), $V^{++}(z,u) = V^{++A} t^A$, where $t^A$ are the generators
of the fundamental representation of the gauge group $G$, normalized by the condition

\begin{equation}
\mbox{tr}(t^A t^B) = \frac{1}{2}\delta^{AB}.
\end{equation}

\noindent
In our notation the structure constants $f^{ABC}$ are defined by the commutation relation

\begin{equation}
[t^A,t^B] = if^{ABC} t^C.
\end{equation}

\noindent
The bare coupling constant $f_0$ in $6D$ has the dimension $m^{-1}$.

Hypermultiplets are described by the analytic superfields $(q^+)_i$, and the covariant harmonic derivative is written as

\begin{equation}
\nabla^{++} = D^{++} + i V^{++} = D^{++} + i V^{++A} T^A\,, \label{Cov++}
\end{equation}

\noindent
where $(T^A)_i{}^j$ denotes generators in the representation $R$ to which the hypermultiplet belongs.

The gauge transformations in harmonic superspace,

\begin{equation}\label{Gauge_Transformations}
V^{++} \to  e^{i\lambda} V^{++} e^{-i\lambda}  - i e^{i\lambda} D^{++}e^{-i\lambda}; \qquad  q^+ \to  e^{i\lambda} q^+,
\end{equation}

\noindent
are parameterized by the Lie-algebra valued analytic superfield parameter $\lambda$.

For quantization we use the background field method
\cite{Buchbinder:2016url,Buchbinder:2017ozh}. Its basic convenience is the
possibility to make the effective action manifestly invariant under the background gauge
transformations by a proper choice of the gauge condition. The background-quantum
splitting is linear,

\begin{equation}
V^{++} = \bm{V}^{++} + v^{++}.
\end{equation}

\noindent
Here $\bm{V}^{++}$ is the background gauge superfield, and $v^{++}$ is the quantum gauge superfield. We will use the
gauge-fixing term in the form

\begin{eqnarray}\label{Gauge_Fixing_Term}
&& S_{\mbox{\scriptsize gf}} = - \frac{1}{2f_0^2}\mbox{tr}\int d^{14}z\, du_1 du_2 \frac{(u_1^- u_2^-)}{(u_1^+ u_2^+)^3} D_1^{++} \Big(e^{-i\bm{b}(z,u_1)}
v^{++}(z,u_1) e^{i\bm{b}(z,u_1)}\Big) \nonumber\\
&& \times D_2^{++} \Big(e^{-i\bm{b}(z,u_2)}v^{++}(z,u_2)e^{i\bm{b}(z,u_2)}\Big)\,,
\end{eqnarray}

\noindent where $\bm{b}(z,u)$ is the background bridge superfield
related to the background superfield $\bm{V}^{++}$ by the relation

\begin{equation}
\bm{V}^{++} = - i e^{i\bm{b}} D^{++} e^{-i\bm{b}}.
\end{equation}

\noindent
The gauge-fixing term (\ref{Gauge_Fixing_Term}) is invariant under the background gauge transformations

\begin{equation}\label{Background_Gauge_Transformations}
\bm{V}^{++} \to  e^{i\lambda} \bm{V}^{++} e^{-i\lambda}  - i e^{i\lambda} D^{++}e^{-i\lambda};
\; v^{++} \to e^{i\lambda} v^{++} e^{-i\lambda};\;   q^+ \to  e^{i\lambda} q^+;\; e^{i\bm{b}}\,,
\to e^{i\lambda} e^{i\bm{b}} e^{i\tau},
\end{equation}

\noindent
where the gauge parameter $\tau=\tau(z)$ does not depend on the harmonic variables.

The Faddeev--Popov ghost action corresponding to the gauge fixing-term (\ref{Gauge_Fixing_Term}) reads

\begin{equation}\label{Action_Faddeev_Popov}
S_{\mbox{\scriptsize FP}} = \mbox{tr} \int d\zeta^{(-4)}\, du\, b \bm{\nabla}^{++}\Big( \bm{\nabla}^{++} c + i[v^{++},c]\Big),
\end{equation}

\noindent where $b$ and $c$ are anticommuting analytic superfields
in the adjoint representation of the gauge group, and
$\bm{\nabla}^{++} c = D^{++} c + i [\bm{V}^{++}, c]$ is the
background-covariant derivative. The generating functional also involves determinants corresponding to the Nielsen--Kallosh
ghosts,

\begin{equation}\label{Z_functional}
Z = \int Dv^{++}\,D\widetilde q^+\, Dq^+\,Db\,Dc\,D\varphi\,\mbox{Det}^{1/2} \stackrel{\bm{\frown}}{\bm{\Box}}
\exp\Big[i(S+S_{\mbox{\scriptsize gf}}
+S_{\mbox{\scriptsize FP}} + S_{\mbox{\scriptsize NK}}+ S_{\mbox{\scriptsize sources}})\Big],
\end{equation}

\noindent
where

\begin{equation}
S_{\mbox{\scriptsize NK}} = -\frac{1}{2}\mbox{tr} \int d\zeta^{(-4)}\, du\, (\bm{\nabla}^{++}\varphi)^2,
\end{equation}

\noindent and $\varphi$ is a commuting analytic superfield in the
adjoint representation. In our notation

\begin{equation}
\stackrel{\bm{\frown}}
{\bm{\Box}}\equiv \frac{1}{2} (D^+)^4 (\bm{\nabla}^{--})^2 \qquad \mbox{and}\qquad (D^+)^4 = - \frac{1}{24}\varepsilon^{abcd} D^+_a D^+_b D^+_c D^+_d.
\end{equation}

\noindent The operator $\stackrel{\bm{\frown}}{\bm{\Box}}$ acts  on
a superfield $\sigma$ in the adjoint representation, for which

\begin{equation}
\bm{\nabla}^{--}\sigma \equiv D^{--}\sigma + i [\bm{V}^{--},\sigma],
\end{equation}

\noindent
where $\bm{V}^{--} = - i e^{i\bm{b}} D^{--} e^{-i\bm{b}}$. The expression $S_{\mbox{\scriptsize sources}}$ includes the relevant source terms.

The structure of divergences in the hypermultiplet
sector is determined by the expression for the superficial degree of divergence in harmonic
superfield formulation \cite{Buchbinder:2016url,Buchbinder:2017ozh}.

\begin{equation}\label{Divergence_Degree}
\omega = 2L - N_q - \frac{1}{2} N_D,
\end{equation}

\noindent where $L$ is a number of loops, $N_q$ is a number of
external hypermultiplet legs, and $N_D$ is a number of spinor
derivatives acting on the external legs. From this equation we see
that in the two-loop approximation ($L=2$) the diagrams with two
external hypermultiplet legs ($N_q=2$) are quadratically divergent.
Also we see that the diagrams with four external hypermultiplet
legs ($N_q=4$) are logarithmically divergent. In this paper we will
calculate the two-point function of the hypermultiplet which
corresponds to the first case only.

\section{Two-loop two-point Green function of the hypermultiplet}
\hspace*{\parindent}

In the one-loop approximation the two-point function of the
hypermultiplet is given by the first diagram in Fig.
\ref{Figure_2Loop}. All other diagrams correspond to two loops. The
last diagram (5) in Fig. \ref{Figure_2Loop} contains the insertion of the
one-loop polarization operator, which is denoted by the gray disk.
The diagrams contributing to this one-loop polarization operator are
depicted in Fig. \ref{Figure_Effective}. Although our purpose is to
calculate these diagrams for ${\cal N}=(1,1)$ theory, we will
consider a more general case of ${\cal N}=(1,0)$ theory with
hypermultiplets in the representation $R$. The direct calculation leads to the following contributions to the effective
action coming from the diagrams drawn in Fig.
\ref{Figure_2Loop} \footnote{The calculations are similar to those
in \cite{Buchbinder:2017ozh} and here we omit the technical details.}:

\begin{figure}
\begin{picture}(0,3)
\put(2.5,2.3){\includegraphics[scale=0.17]{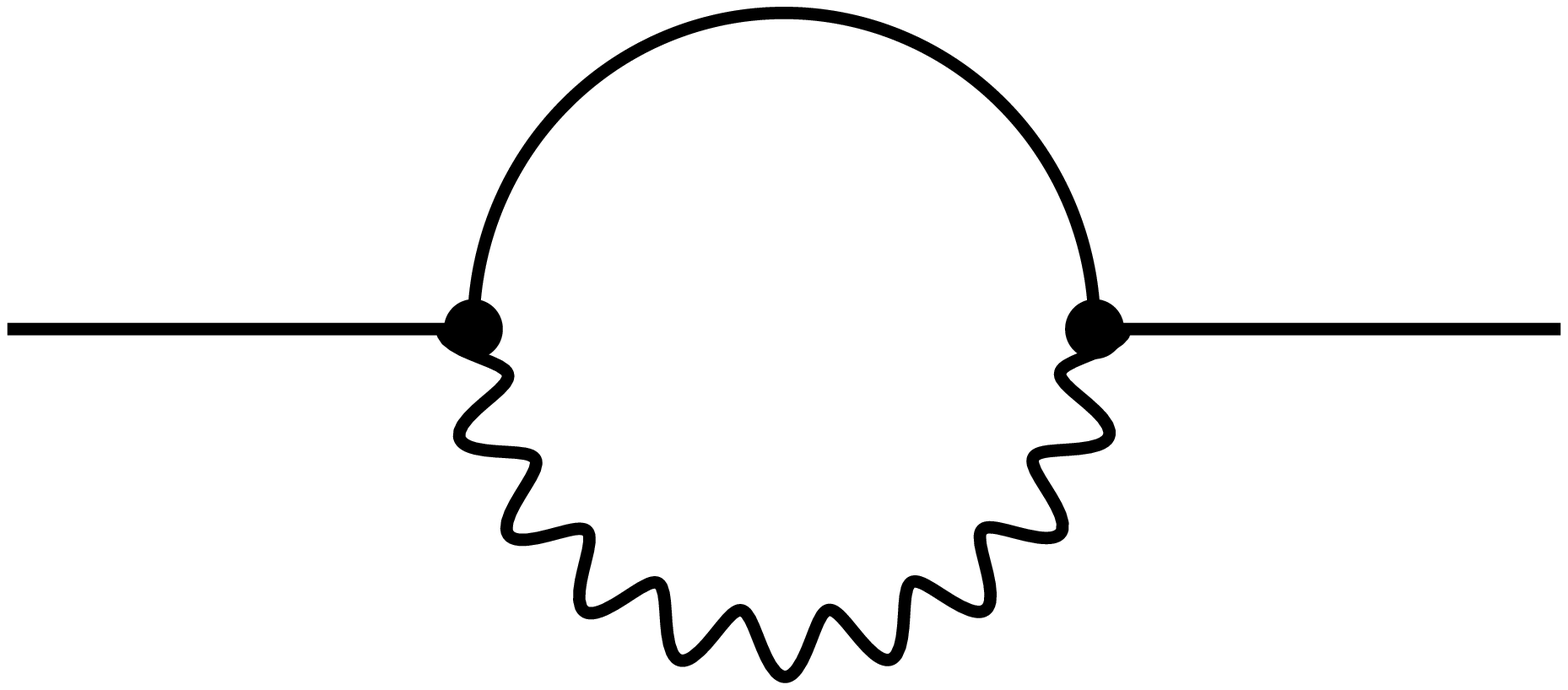}}
\put(6.5,2.3){\includegraphics[scale=0.17]{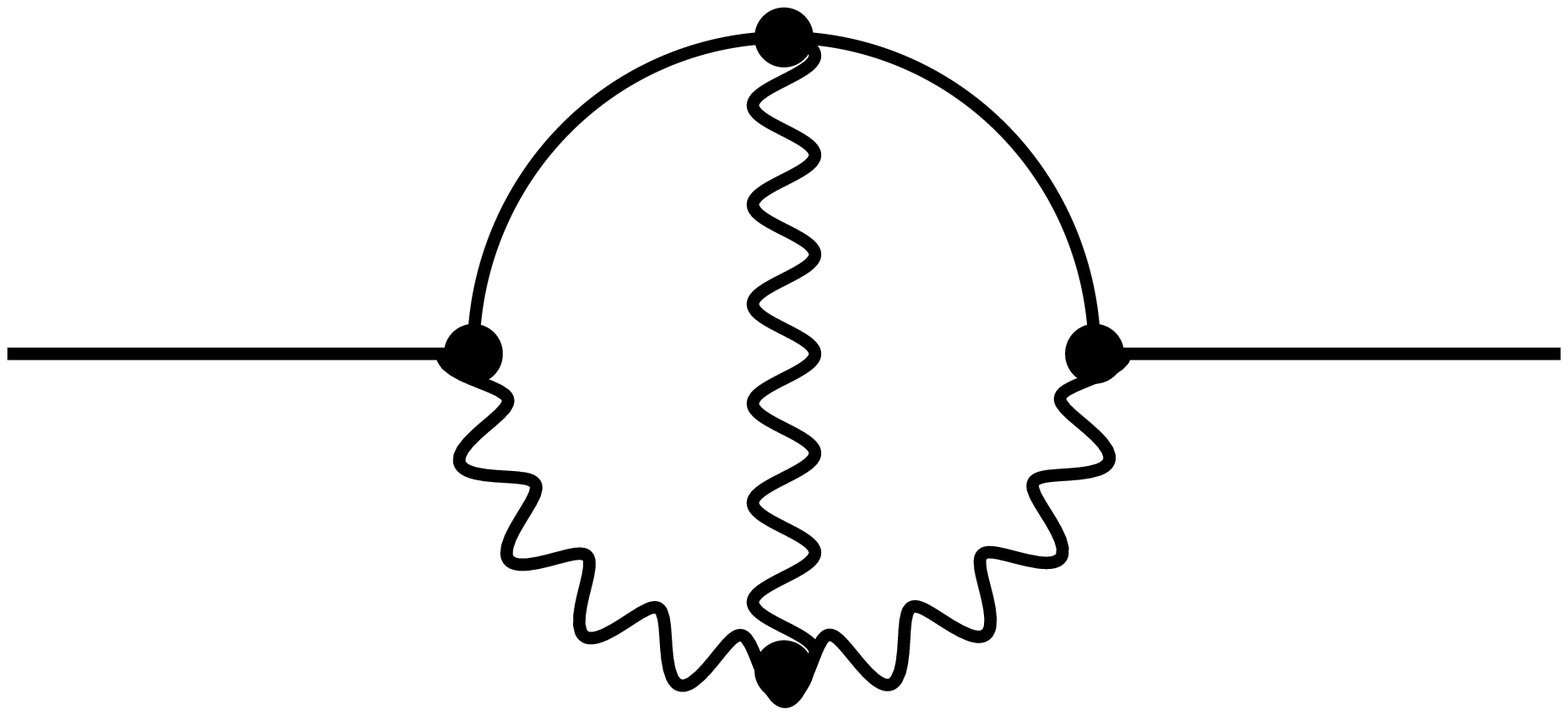}}
\put(10.5,2.3){\includegraphics[scale=0.17]{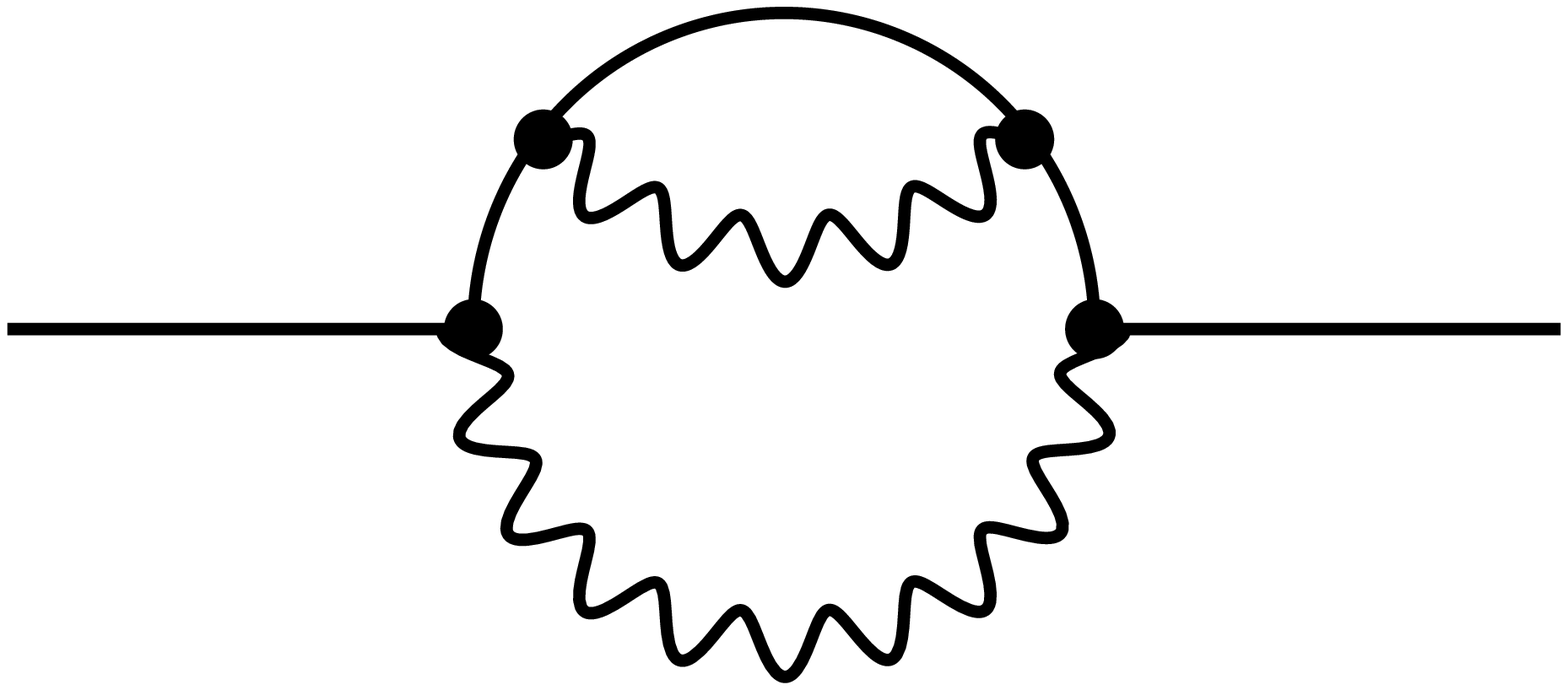}}
\put(4.5,0){\includegraphics[scale=0.17]{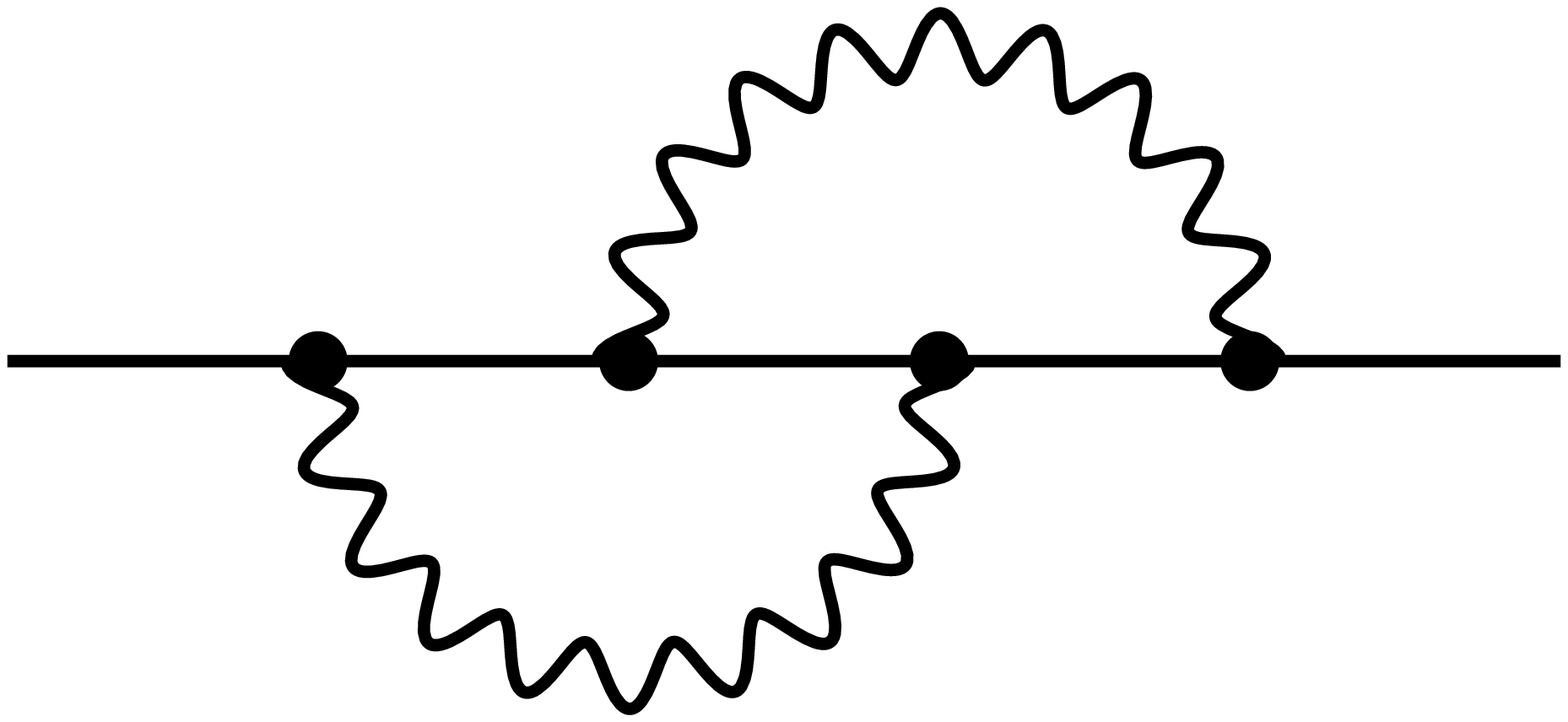}}
\put(8.5,-0.1){\includegraphics[scale=0.17]{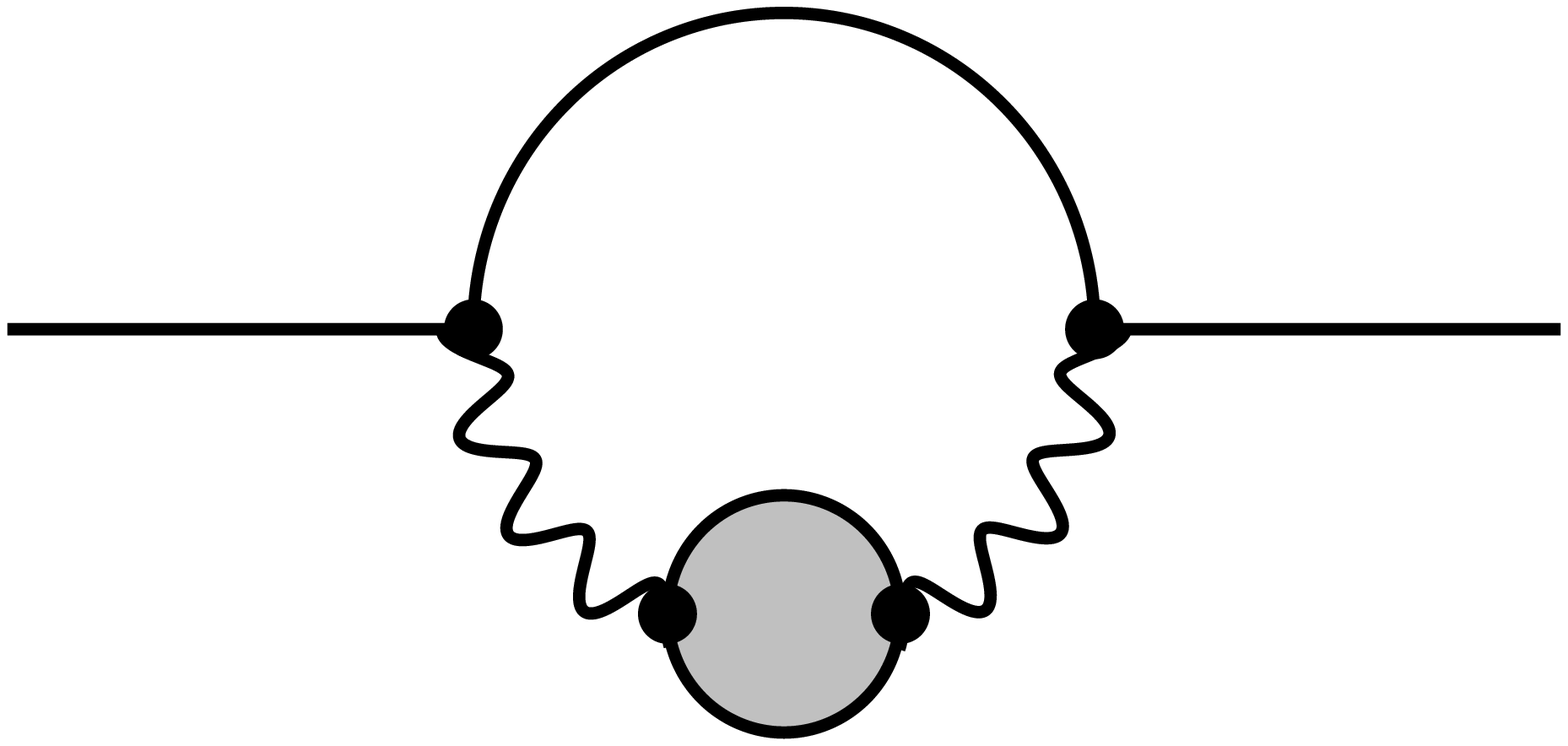}}
\put(2.4,3.7){$(1)$}
\put(6.4,3.7){$(2)$}
\put(10.5,3.7){$(3)$}
\put(4.5,1.3){$(4)$}
\put(8.5,1.3){$(5)$}
\end{picture}
\caption{One- and two-loop diagrams contributing to the two-point Green function of the hypermultiplet.}\label{Figure_2Loop}
\end{figure}

\begin{figure}
\begin{picture}(0,4)
\put(2.2,2.0){\includegraphics[scale=0.17]{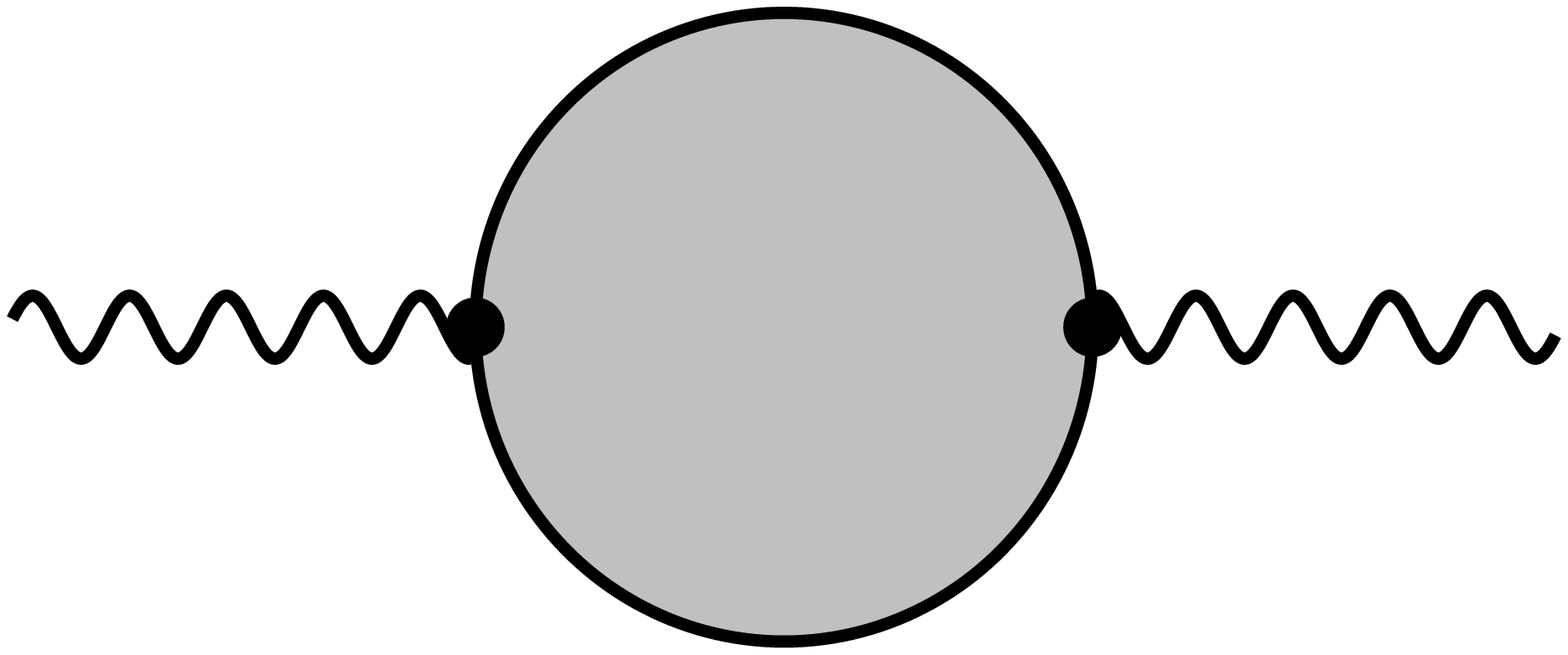}}
\put(5.9,2.6){$=$}
\put(10.5,2.0){\includegraphics[scale=0.4]{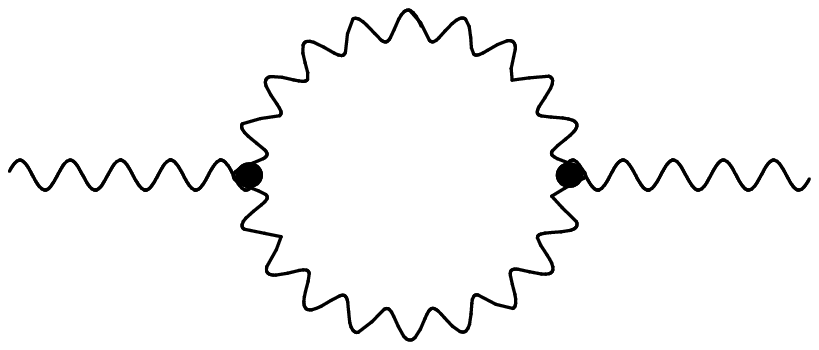}}
\put(9.6,2.6){$+$}
\put(6.7,1.7){\includegraphics[scale=0.4]{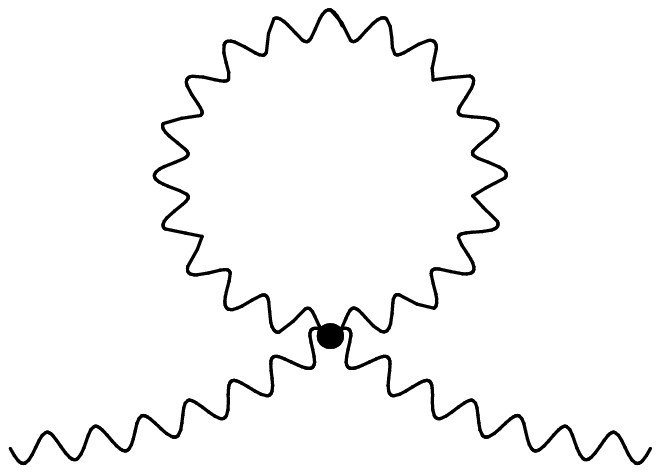}}
\put(6.2,-0.01){\includegraphics[scale=0.4]{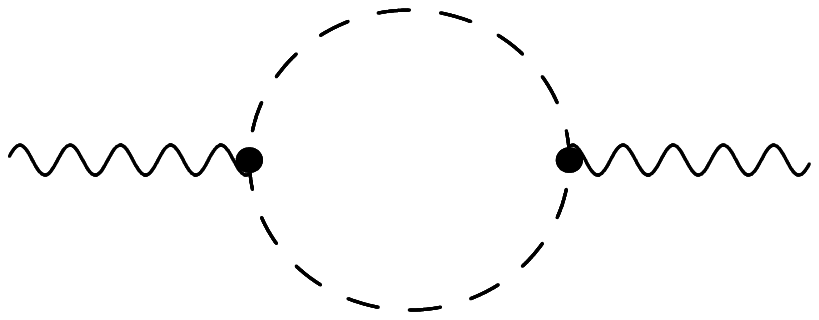}}
\put(10.6,-0.02){\includegraphics[scale=0.4]{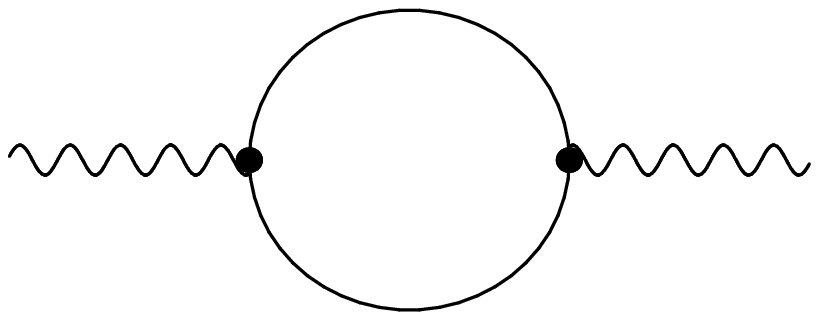}}
\put(5.5,0.5){$+$} \put(10.0,0.5){$+$}
\end{picture}
\caption{One-loop subdiagrams entering the two-loop diagram (5) in Fig. \ref{Figure_2Loop}.}\label{Figure_Effective}
\end{figure}

\begin{eqnarray}
&& (1) = 0;\vphantom{\frac{1}{2}}\\
&& (2) = 2 C_2 f_0^4 \int \frac{d^6p}{(2\pi)^6} d^8\theta\, \int\frac{du_1\, du_2}{(u_1^+ u_2^+)}\, \widetilde q^+(p,\theta,u_1)^i (T^A T^A)_i{}^j q^+(-p,\theta,u_2)_j\nonumber\\
&& \times \int \frac{d^6k}{(2\pi)^6} \frac{d^6l}{(2\pi)^6} \frac{1}{k^2 l^2 (k+l)^2 (k+l+p)^2 (k+p)^2};\\
&& (3) = 0;\vphantom{\frac{1}{2}}\\
&& (4) = -4 f_0^4 \int \frac{d^6p}{(2\pi)^6} d^8\theta\, \int\frac{du_1\, du_2}{(u_1^+ u_2^+)}\, \widetilde q^+(p,\theta,u_1)^i (T^A T^B T^A T^B)_i{}^j q^+(-p,\theta,u_2)_j\nonumber\\
&& \times \int \frac{d^6k}{(2\pi)^6} \frac{d^6l}{(2\pi)^6} \frac{1}{k^2 l^2 (k+l)^2 (k+l+p)^2 (k+p)^2};\\
&& (5)= 4 \Big(C_2 - T(R)\Big) f_0^4 \int \frac{d^6p}{(2\pi)^6} d^8\theta\, \int\frac{du_1\, du_2}{(u_1^+ u_2^+)}\, \widetilde q^+(p,\theta,u_1)^i (T^A T^A)_i{}^j q^+(-p,\theta,u_2)_j\nonumber\\
&& \times \int \frac{d^6k}{(2\pi)^6} \frac{d^6l}{(2\pi)^6} \frac{1}{k^4 (k+p)^2 l^2 (k+l)^2}.
\end{eqnarray}

\noindent
Note that all these expressions are written in the
Minkowski space before the Wick rotation. The group theory
coefficients entering them are defined by the relations

\begin{equation}
f^{ACD} f^{BCD} = C_2\delta^{AB};\qquad T(R)\delta^{AB} = \mbox{tr}(T^A T^B);\qquad C(R)_i{}^j = (T^A T^A)_i{}^j.
\end{equation}

\noindent {}From these relations,  after some algebra, we derive

\begin{equation}
(T^A T^B T^A T^B)_i{}^j = \Big(C(R)^2 - \frac{1}{2} C_2 C(R)\Big)_i{}^j.
\end{equation}

\noindent Thus, the result for the sum of all considered
diagrams can be written in the form

\begin{eqnarray}\label{Final_Result}
&& 4 f_0^4 \int \frac{d^6p}{(2\pi)^6} d^8\theta\, \int\frac{du_1\, du_2}{(u_1^+ u_2^+)}\, \Bigg[\widetilde q^+(p,\theta,u_1)^i \Big(-C(R)^2 + C_2 C(R)\Big)_i{}^j q^+(-p,\theta,u_2)_j\nonumber\\
&&\times \int \frac{d^6k}{(2\pi)^6} \frac{d^6l}{(2\pi)^6} \frac{1}{k^2 l^2 (k+l)^2 (k+l+p)^2 (k+p)^2}
+ \Big(C_2-T(R)\Big)\widetilde q^+(p,\theta,u_1)^i  C(R)_i{}^j\qquad\nonumber\\
&& \times q^+(-p,\theta,u_2)_j \int \frac{d^6k}{(2\pi)^6} \frac{d^6l}{(2\pi)^6} \frac{1}{k^4 (k+p)^2 l^2 (k+l)^2} \Bigg].
\end{eqnarray}

\noindent We see that it is quadratically divergent, in the precise
agreement with the general expression for the degree of divergence
calculated in \cite{Buchbinder:2016url}. The expression
(\ref{Final_Result}) is written formally, because the regularization
was not still introduced. It is known that quadratic divergences cannot be catched within the dimensional regularization,
and it is necessary to use different regularization schemes,  as, e.g., in  \cite{Buchbinder:2015eva}. Note, however, that for the  considered theory $R=Adj$,
i.e. the hypermultiplets belong to the adjoint representation of the gauge group,

\begin{equation}\label{condition}
T(Adj)=C_2;\qquad C(Adj)_i{}^j = C_2 \delta_i^j.
\end{equation}

\noindent This implies that for ${\cal N}=(1,1)$ theory the
expression (\ref{Final_Result}) vanishes identically. In particular,
the leading quadratic divergences cancel each other, so that in the
case of using the dimensional regularization technique we do not
miss any divergent contributions.  Moreover, within the dimensional
reduction scheme the result obtained from Eq. (\ref{Final_Result}) by the
replacement $6\to D$ also identically vanishes. Thereby, the
logarithmically divergent contributions are also absent and the
considered Green function vanishes in the two-loop approximation for
${\cal N}=(1,1)$ SYM theory.

It is worth to  point out that the
finiteness of the two-point supergraphs in the hypermultiplet sector,
implied by the conditions (\ref{condition}), is achieved in the same way as in
the one-loop case \cite{Buchbinder:2016url,Buchbinder:2017ozh}.

\section*{Summary}
\hspace*{\parindent}

In this paper we have investigated the two-loop divergences in
$6D,\,{\cal N}=(1,1)$ SYM theory. This theory is a $6$-dimensional
analog of $4D, \,{\cal N}=4$ SYM theory, for which reason one could expect a better
ultraviolet behavior of this theory in comparison with other $6D$ theories.
First, we calculated the two-loop
divergences of the hypermultiplet two-point function in $6D, {\cal
N}=(1,0)$ vector multiplet theory coupled to the hypermultiplet in
an arbitrary representation of gauge group. Then we turn to
$6D,\, {\cal N}=(1,1)$ SYM theory, which corresponds to the hypermultiplet in
adjoint representation. We proved that the corresponding
divergences identically vanish\footnote{Here we essentially used
the property that off-shell divergences in the theory under consideration are
absent at one loop \cite{Buchbinder:2016url,Buchbinder:2017ozh}
and therefore there is no need to take into account the one-loop counterterms
for two-loop diagrams.} without using the equations of motion.
Moreover, the conditions of vanishing of divergences are
the same as in the one-loop case. Taking into account that the
Green function considered is related to other two-point Green
functions of ${\cal N}=(1,1)$ theory by hidden ${\cal N}=(0,1)$ supersymmetry,
we come to the conclusion that all two-point Green functions of the theory are finite in the two-loop
approximation. However, logarithmic divergences can still appear in
the four-point Green functions. To see, whether they are
finite or not, it will be sufficient to calculate the four-point function of the
hypermultiplet. We are going to address this problem in the forthcoming paper.

\subsection*{Acknowledgments }

The work of K.V.S was supported by Russian Scientific Foundation, project No. 16-12-1036. The work of I.L.B., E.A.I. and B.M.M. was partially supported by Russian Scientific Foundation, project No. 16-12-1036, RFBR, project No. 15-02-06670 and RFBR-DFG, project No. 16-52-12012.


\begin{thebibliography}{99}

\bibitem{Buchbinder:2016gmc}
  I.~L.~Buchbinder, E.~A.~Ivanov, B.~S.~Merzlikin and K.~V.~Stepanyantz,
  ``One-loop divergences in the 6D, N=(1,0) abelian gauge theory'',
  Phys. Lett. B {\bf 763} (2016) 375-381,
  [arXiv:1609.00975 [hep-th]].

\bibitem{Buchbinder:2016url}
  I.~L.~Buchbinder, E.~A.~Ivanov, B.~S.~Merzlikin and K.~V.~Stepanyantz,
``One-loop divergences in 6D, N=(1,0) SYM theory'',
  JHEP {\bf 1701} (2017) 128,
  [arXiv:1612.03190 [hep-th]].

\bibitem{Buchbinder:2017ozh}
  I.~L.~Buchbinder, E.~A.~Ivanov, B.~S.~Merzlikin and K.~V.~Stepanyantz,
  ``Supergraph analysis of the one-loop divergences in $6D$, ${\cal N} = (1,0)$ and ${\cal N} = (1,1)$ gauge theories,''
  Nucl.\ Phys.\ B {\bf 921} (2017) 127,
  [arXiv:1704.02530 [hep-th]].

\bibitem{Howe:1983jm}
  P.~S.~Howe and K.~S.~Stelle,
  ``Ultraviolet divergences in higher dimensional supersymmetric {Yang-Mills} theories'',
  Phys.\ Lett.\ B {\bf 137} (1984) 175-180.

\bibitem{Howe:2002ui}
  P.~S.~Howe and K.~S.~Stelle,
  ``Supersymmetry counterterms revisited'',
  Phys.\ Lett.\ B {\bf 554} (2003) 190-196,
  [arXiv:hep-th/0211279].

\bibitem{Bossard:2009sy}
  G.~Bossard, P.~S.~Howe and K.~S.~Stelle,
  ``The ultra-violet question in maximally supersymmetric field theories'',
  Gen.\ Rel.\ Grav.\  {\bf 41} (2009) 919-981,
  [arXiv:0901.4661 [hep-th]].

\bibitem{Bossard:2009mn}
  G.~Bossard, P.~S.~Howe and K.~S.~Stelle,
  ``A note on the UV behaviour of maximally supersymmetric Yang-Mills theories'',
  Phys.\ Lett.\ B {\bf 682} (2009) 137-142,
  [arXiv:0908.3883 [hep-th]].

\bibitem{Fradkin:1982kf}
  E.~S.~Fradkin and A.~A.~Tseytlin,
  ``Quantum properties of higher dimensional and dimensionally reduced supersymmetric theories'',
  Nucl.\ Phys.\ B {\bf 227} (1983) 252-290.

\bibitem{Marcus:1983bd}
  N.~Marcus and A.~Sagnotti,
  Phys.\ Lett.\  {\bf 135B} (1984) 85.

\bibitem{Marcus:1984ei}
  N.~Marcus and A.~Sagnotti,
  ``The Ultraviolet Behavior of $N=4$ {Yang-Mills} and the Power Counting of Extended Superspace,''
  Nucl.\ Phys.\ B {\bf 256} (1985) 77.

\bibitem{Bork:2015zaa}
  L.~V.~Bork, D.~I.~Kazakov, M.~V.~Kompaniets, D.~M.~Tolkachev and D.~E.~Vlasenko,
  ``Divergences in maximal supersymmetric Yang-Mills theories in diverse dimensions'',
  JHEP {\bf 1511} (2015) 059,
  [arXiv:1508.05570 [hep-th]].
\bibitem{Bossard:2015dva}
  G.~Bossard, E.~Ivanov and A.~Smilga,
  ``Ultraviolet behavior of 6D supersymmetric Yang-Mills theories and harmonic superspace'',
  JHEP {\bf 1512} (2015) 085,
  [arXiv:1509.08027 [hep-th]].


\bibitem{Smilga:2016dpe}
  A.~Smilga,
  ``Ultraviolet divergences in non-renormalizable supersymmetric theories,''
  Phys.\ Part.\ Nucl.\ Lett.\  {\bf 14} (2017) no.2,  245,
  [arXiv:1603.06811 [hep-th]].

\bibitem{Grisaru:1982zh}
  M.~T.~Grisaru and W.~Siegel,
  Nucl.\ Phys.\ B {\bf 201} (1982) 292
   [Erratum-ibid.\ B {\bf 206} (1982) 496].

\bibitem{Mandelstam:1982cb}
  S.~Mandelstam,
  Nucl.\ Phys.\ B {\bf 213} (1983) 149.

\bibitem{Brink:1982pd}
  L.~Brink, O.~Lindgren and B.~E.~W.~Nilsson,
  Nucl.\ Phys.\ B {\bf 212} (1983) 401.

\bibitem{Howe:1983sr}
  P.~S.~Howe, K.~S.~Stelle and P.~K.~Townsend,
  Nucl.\ Phys.\ B {\bf 236} (1984) 125.

\bibitem{Howe:1983fr}
  P.~S.~Howe, G.~Sierra and P.~K.~Townsend,
  ``Supersymmetry in six-dimensions'',
  Nucl.\ Phys.\ B {\bf 221} (1983) 331-348.

\bibitem{Howe:1985ar}
  P.~S.~Howe, K.~S.~Stelle and P.~C.~West,
  ``N=1, d = 6 harmonic superspace'',
  Class.\ Quant.\ Grav.\  {\bf 2} (1985) 815-821.

\bibitem{Zupnik:1986da}
  B.~M.~Zupnik,
  ``Six-dimensional supergauge theories in the harmonic superspace'',
  Sov.\ J.\ Nucl.\ Phys.\  {\bf 44} (1986) 512
   [Yad.\ Fiz.\  {\bf 44} (1986) 794-802].

\bibitem{Ivanov:2005qf}
  E.~A.~Ivanov, A.~V.~Smilga and B.~M.~Zupnik,
  ``Renormalizable supersymmetric gauge theory in six dimensions'',
  Nucl.\ Phys.\ B {\bf 726} (2005) 131-148,
  [arXiv:hep-th/0505082].

\bibitem{Ivanov:2005kz}
  E.~A.~Ivanov and A.~V.~Smilga,
  ``Conformal properties of hypermultiplet actions in six dimensions'',
  Phys.\ Lett.\ B {\bf 637} (2006) 374-381,
  [arXiv:hep-th/0510273].

\bibitem{Buchbinder:2014sna}
  I.~L.~Buchbinder and N.~G.~Pletnev,
  ``Construction of 6D supersymmetric field models in N=(1,0) harmonic superspace'',
  Nucl.\ Phys.\ B {\bf 892} (2015) 21-48,
  [arXiv:1411.1848 [hep-th]].


\bibitem{Townsend:1983ana}
  P.~K.~Townsend and G.~Sierra,
  ``Chiral anomalies and constraints on the gauge group in higher dimensional supersymmetric {Yang-Mills} theories'',
  Nucl.\ Phys.\ B {\bf 222} (1983) 493-506.

\bibitem{Smilga:2006ax}
  A.~V.~Smilga,
  ``Chiral anomalies in higher-derivative supersymmetric 6D theories'',
  Phys.\ Lett.\ B {\bf 647} (2007) 298-304,
  [arXiv:hep-th/0606139].

\bibitem{Kuzenko:2015xiz}
  S.~M.~Kuzenko, J.~Novak and I.~B.~Samsonov,
  ``The anomalous current multiplet in 6D minimal supersymmetry'',
  JHEP {\bf 1602} (2016) 132,
  [arXiv:1511.06582 [hep-th]].

\bibitem{Buchbinder:2015eva}
  I.~L.~Buchbinder, N.~G.~Pletnev and K.~V.~Stepanyantz,
  ``Manifestly N=2 supersymmetric regularization for N=2 supersymmetric field theories,''
  Phys.\ Lett.\ B {\bf 751} (2015) 434,
  [arXiv:1509.08055 [hep-th]].

\end{thebibliography}
\end{document}